\documentclass[11pt]{article}

\let\ssection=\section
\renewcommand{\section}{\setcounter{equation}{0}\ssection}

\def\p{{\partial}}

\def\vnabla{{\mathbf{\nabla}}}

\newcommand{\br}{{\bf r}}
\newcommand{\bv}{{\bf v}}
\newcommand{\bk}{{\bf k}}
\newcommand{\bP}{{\bf P}}
\newcommand{\bQ}{{\bf Q}}
\newcommand{\bE}{{\bf E}}
\newcommand{\bB}{{\bf B}}
\newcommand{\vOmega}{{\mathbf{\Omega}}}
\newcommand{\half}{{\scriptstyle{\frac{1}{2}}}}
\def\det{{\rm det}}
\def\cA{{\cal A}}

\begin{document}

\setlength{\baselineskip}{16pt}

\title{Berry phase correction to electron density  in solids 
 and ``exotic'' dynamics} 

\author{
C. DUVAL\footnote{mailto: duval@cpt.univ-mrs.fr}\\
Centre de Physique Th\'eorique, CNRS, 
Luminy, Case 907\\ 
F-13288 Marseille Cedex 9 (France)\footnote{ 
UMR 6207 du CNRS associ\'ee aux 
Universit\'es d'Aix-Marseille I et II et Universit\'e du Sud Toulon-Var; Laboratoire 
affili\'e \`a la FRUMAM-FR2291
}
\\
Centre de Physique Th\'eorique, CNRS\\
Luminy, Case 907\\
F-13 288 MARSEILLE Cedex 9 (France)
\\
Z.~Horv\'ath\footnote{e-mail: zalanh@ludens.elte.hu}
\\
Institute for Theoretical Physics, E\"otv\"os
University\\
P\'azm\'any P. s\'et\'any 1/A\\
H-1117 BUDAPEST (Hungary)
\\
P.~A.~Horv\'athy\footnote{e-mail: horvathy@lmpt.univ-tours.fr}
\\
Laboratoire de Math\'ematiques et de Physique Th\'eorique\\
Universit\'e de Tours\\
Parc de Grandmont\\
F-37 200 TOURS (France)
\\
L. Martina\footnote{e-mail: Luigi.Martina@le.infn.it}
\\
Dipartimento di Fisica dell'Universit\`a
\\
and\\
Sezione INFN di Lecce. Via Arnesano, CP. 193\\
I-73 100 LECCE (Italy).
\\ and\\
P.~C.~Stichel\footnote{e-mail: peter@Physik.Uni-Bielefeld.DE}
\\
An der Krebskuhle 21\\
D-33 619 BIELEFELD (Germany)
}

\date{\today}

\maketitle

\begin{abstract}
Recent results on the semiclassical dynamics of an electron in a solid
are explained using techniques developed for ``exotic'' Galilean dynamics.   
The system is indeed Hamiltonian and 
Liouville's theorem holds for the symplectic volume form.
Suitably defined quantities satisfy hydrodynamic equations.
\end{abstract}

\vskip0.5mm\noindent
\texttt{cond-mat/0506051}.  
to appear in {\sl Mod. Phys. Lett}. {\bf B}.

\newpage


The effective semiclassical dynamics of a Bloch electron in a solid 
is modified by a Berry curvature term \cite{Bloch}. 
Putting $\hbar=1$, the equations of motion in the $n{}^{th}$ band read indeed
\begin{eqnarray}
\dot{\br}&=\displaystyle\frac{\p\epsilon_n(\bk)}{\p\bk}-\dot{\bk}\times\vOmega(\bk),\label{velrel}
\\[6pt]
\dot{\bk}&=-e\bE-e\dot{\br}\times\bB(\br),
\label{Lorentz}
\end{eqnarray}
where $\br=(x^i)$ and $\bk=(k_j)$ denote the electron's intracell position and quasimomentum,
respectively, $\epsilon_n(\bk)$ is the band energy. The purely momentum-dependent 
$\vOmega$ is the Berry curvature of the electronic Bloch states,
$\Omega_i(\bk)=\epsilon_{ijl}\p_{\bk_j}\cA_l(\bk)$, where 
$\cA$ is the Berry connection. 
The electronic charge is $-e$. The new Berry-curvature dependent term in eqn. (\ref{velrel})
has been instrumental in explaining the anomalous Hall effect in ferromagnetic materials \cite{AHE}, and also the spin Hall effect \cite{spinHall}.

Much of the transport phenomena rely on the 
Liouville's theorem on the conservation of the
phase-space volume element.
Recently, \cite{XSN}  Xiao et al. observed that, due to
the Berry curvature term, the naive phase-space volume form $d\br d\bk$ is no more conserved. They
attribute this fact to the non-Hamiltonian character
of the equations (\ref{velrel})-(\ref{Lorentz}), and
suggest to redefine the phase-space density by including a pre-factor,
\begin{equation}
\rho\to \rho_n=D_nf_n(\br,\bk,t),\qquad
D=\frac{1}{(2\pi)^d}(1+e\bB\cdot\vOmega),
\label{XSNdens}
\end{equation}
where $d$ is the spatial dimension and $f_n$ is the occupation number of the state labeled by $\br,\bk$.  If  the latter is conserved along
the motion, $df_n/dt=0$, then $\rho_n$ satisfies the continuity equation on phase space. Eqn. (\ref{XSNdens}) is the starting point of Xiao et al. to derive interesting applications.

Xiao et al.  derive Eq. (\ref{XSNdens}), their main result, by calculating
the change of $d\br d\bk$ using the equations of motion and then finding a compensating factor. This may appear somewhat mysterious at first sight.
Our Note is to the end to point out that
the equations (\ref{velrel})-(\ref{Lorentz}) {\it are} 
indeed Hamiltonian in a by-now standard sense \cite{Hamdyn}, and 
 the validity of Liouville's theorem is restored if the symplectic volume form, Eq. (\ref{volel}) below, is used. Then Eq. (\ref{XSNdens}) follows at once. Our clue is the relation to ``exotic'' Galilean dynamics, introduced independently and equivalent to non-commutative mechanics in  \cite{LSZ,DH,DHH,HMS}.

Let us first summarize the general principles of Hamiltonian dynamics \cite{Hamdyn}.
The first ingredient is the symplectic form on $2d$-dimensional phase space,
i.e., a closed and regular 2-form 
$
\omega=\half\omega_{\alpha\beta}d\xi^\alpha\wedge d\xi^\beta,
$ 
where $\xi^\alpha$ denotes the
phase space coordinates $\br$ and $\bk$ collectively. Then the Poisson bracket of two functions $f$ and $g$ on phase space is  
 \begin{equation}
 \{f,g\}=\omega^{\alpha\beta}\p_\alpha f\p_\beta g,
 \label{PB}
 \end{equation}
 where $\omega^{\alpha\beta}$ is the {\it inverse} of the symplectic matrix,
 $\omega^{\alpha\gamma}\omega_{\gamma\beta}=\delta^{\alpha}_{\ \beta}$.
Since $\omega$ is closed, the Poisson bracket (\ref{PB}) satisfies the Jacobi identity. Then Hamilton's equations read 
\begin{equation}
 \dot{\xi}^{\alpha}=\{h,\xi^{\alpha}\},
 \label{hameq}
\end{equation} 
where $h=h(\xi)$ is some given Hamiltonian, which is the second ingredient we need.
Such a framework is obtained \cite{FaJa}, in particular, starting with a first-order Lagrangian 
$L=a_\alpha(\xi)\dot{\xi}^\alpha-h(\xi)$ on phase-space,
whose  Euler-Lagrange equations read
 \begin{equation}
 \omega_{\alpha\beta}\dot{\xi}^\beta=\p_\alpha h,
 \quad\hbox{where}\quad
 \omega_{\alpha\beta}=\p_\alpha a_\beta-\p_\beta a_\alpha,
 \label{EL}
 \end{equation}
and multiplying (\ref{EL}) by the inverse matrix  $\omega^{\alpha\beta}$ yields  Hamilton's equations, (\ref{hameq}).

\goodbreak
 
Now the natural volume form on phase space is the $d^{th}$
 power of the symplectic form \cite{Hamdyn} which, in {\it arbitrary} coordinates $\xi^{\alpha}$ on phase space, reads
 \begin{equation}
 dV=\frac{1}{d!}\omega^d=\sqrt{\det(\omega_{\alpha\beta})}\prod_{\alpha=1}^{2d}d\xi^\alpha.
 \label{volel}
 \end{equation}

The  intrinsic expression (\ref{volel}) is  invariant
w.r.t. coordinate transformations. The naive expression 
$\prod d\xi^\alpha$, changes in fact as   
$
\prod d\xi^\alpha\to\prod d\zeta^\alpha
=
\det({\p^\alpha\zeta}/{\p\xi^\beta})\prod d\xi^\alpha
$
under a transformation $\xi\to\zeta$. This is, however, compensated by the
change of the determinant. 
 As a bonus, the validity of Liouville's theorem is restored:  the r.h.s. of (\ref{hameq}) generates the classical flow of the phase space,
 w.r.t. which  the
symplectic volume form, (\ref{volel}), {\it is} invariant, since 
 the classical flow is made of symplectic transformations \cite{Hamdyn}.  

Now Darboux's theorem tells that one can always 
find canonical local coordinates 
$p_i, q^j$ (say) such that the symplectic matrix has the canonical form
$
\omega=dp_i\wedge dq^i
$.
 Then the general Hamilton equations, (\ref{hameq}) take their
 familiar form $\dot{q}_i=\p_{p_{i}}h$, $\dot{p}_j=-\p_{q_{j}}h$.
 The volume form is simply $\prod dp_i\wedge dq^i$.
 Xiao et al.  use the restricted terminology, referring to canonical coordinates.  
Non-canonical coordinates  may be useful, though.
The equations of motion (\ref{velrel})-(\ref{Lorentz}) can indeed be put into the form (\ref{EL}) with symplectic and resp. Poisson matrix
\begin{equation}
\omega_{\alpha\beta}=\left(
\begin{array}{cc}
-\varepsilon_{ijk}eB^k&-\delta_{jk}
\\[8pt]
\delta_{jk}&\varepsilon_{ijk}\Omega^k
\end{array}\right)
\quad\Leftrightarrow\quad
\omega^{\alpha\beta}=\frac{1}{1+e\bB\cdot\vOmega}\left(
\begin{array}{cc}
\varepsilon_{ijk}\Omega^k&\delta_{jk}+eB_j\Omega_k
\\[8pt]
-\delta_{jk}-e\Omega_jB_k&-\varepsilon_{ijk}eB^k
\end{array}\right)
\label{gensymp}
\end{equation}
where $i,j,k=1,2,3$. The Poisson bracket relations are therefore  
\vspace{-3mm}
\begin{eqnarray}
\{x^i,x^j\}&=\displaystyle\frac{\varepsilon^{ijk}\Omega_k}{1+e\bB\cdot\vOmega},
\label{xx}
\\[5pt]
\{x^i,k_j\}&=\displaystyle\frac{\delta^{i}_{\ j}+eB^i\Omega_j}{1+e\bB\cdot\vOmega},
\label{kk}
\\[5pt]
\{k_i,k_j\}&=-\displaystyle\frac{\varepsilon_{ijk}eB^k}{1+e\bB\cdot\vOmega}.
\label{xk}
\end{eqnarray}
The Berry curvature is divergence-free, $\p_{\bk_j}\Omega^j=0$,
because it is a curl.
The Jacobi identity follows hence from 
$\p_{x^j}B^j=0$,
cf.  \cite{BM}. 
The Hamiltonian is $h=\epsilon_n-eV$.
Note, in particular, that the coordinates do not commute: the model is indeed {\it equivalent to non-commutative mechanics} \cite{LSZ,DH,BM}.

When both the  $\bB$, and the Berry curvature, $\vOmega=$const. 
are directed in the third direction, these relations reduce to those of the ''exotic'' model in the plane, derived from first principles  \cite{DH,HMS}.
The crucial  anomalous velocity term in
(\ref{velrel}) arises, in particular, by minimal
coupling of the ''exotic'' model to a gauge
field \cite{DH}.
Eq. (\ref{gensymp}) is obtained, furthermore, by the Faddeev-Jackiw \cite{FaJa} construction mentioned above, if we start with the  Lagrangian \cite{Bloch,DH,HMS}
\begin{equation}
     L^{Bloch}=\big(k_{i}-eA_{i}(\br,t)\big)\dot{x}^{i}-
     \big(\epsilon_n(\bk)-eV(\br,t)\big)
     +\cA^{i}(\bk)\dot{k}_{i},
     \label{blochlag}
\end{equation}
where $V$ and $A_{i}$ are the scalar  resp.
 vector potential for the electromagnetic field and  
 $\cA$ is the potential  for the Berry  curvature.
The conserved volume form is \cite{DHH}
 \begin{equation}
 \sqrt{\det(\omega_{\alpha\beta})}\,\prod dk_i\wedge dx^i=({1+e\bB\cdot\vOmega})\prod_{i}
 dk_i\wedge dx^i.
 \label{exovol}
 \end{equation}
Consistently with Darboux's theorem, canonical coordinates can be found.

 In the planar case, for example, if $B$ is constant such  that $1+eB\Omega\neq0$, 
\begin{equation}
\begin{array}{l}
P_{i}=\displaystyle\sqrt{1+eB\Omega}\,k_i
+\frac{1}{2} eB\varepsilon_{ij}Q^{j}\hfill
\\[3.6mm]
Q^{i}=x^{i}-
\displaystyle\frac{1}{eB}\left(1-\sqrt{1+eB\Omega}\right)
\varepsilon^{ij}k_{j}\hfill
\end{array}
\label{cancoord}
\end{equation}
are canonical \cite{DHH}, so that the  Poisson bracket reads simply
$
\big\{F,G\big\}=
\sum\p_{Q_i}F\,\p_{P_i}G
-
\p_{Q_i}G\,\p_{P_i}F.
$ 
(The disadvantage is that the Hamiltonian becomes rather complicated).
When expressed in terms of canonical coordinates,
the four-dimensional  volume form reads simply
$dQ_{1}\wedge dQ_{2}\wedge dP_{1}\wedge dP_{2}$; the
$2D$ version of (\ref{volel}) is recovered as 
$ 
dV={\det\big(\p(\bP,\bQ)/\p(\bk,\br)\big)}\,d\br d\bk.
$
Similar formul{\ae} would work in higher dimensions.

Let us consider a particle distribution $f=f(\br,\bk,t)$. The probability of finding our particle in the phase space volume element $dV$ is, see (\ref{exovol}),
\begin{equation}
fdV=
f\sqrt{\det(\omega_{\alpha\beta})}\prod_\alpha d\xi^\alpha
=
f(\br,\bk,t)\big(1+e\bB\cdot\vOmega\big)dx_1dx_2dx_3dk_1dk_2dk_3.
\label{prob}
\end{equation}
Since $dV=\omega^3/6$ is invariant by the Liouville theorem, $fdV$ is an integral invariant of the flow, provided $df/dt=0$, i. e.,
$\p_tf+\{f,h\}=0$. However, 
$$
\p_tf+\vnabla_{\xi}\cdot(\dot{\xi}f)=-\frac{e(\bB\cdot\vOmega)\dot{\,}}
{1+e\bB\cdot\vOmega}f,
$$
so that $f$ can not be considered as a density.
Multiplication with the square root of the determinant
removes the unwanted term, though~:
\begin{equation}
\rho=f\sqrt{\det(\omega_{\alpha\beta})}
\label{gooddensity}
\end{equation} 
 {\it does} satisfy the continuity equation 
 $\p_t\rho+\vnabla_{\xi}\cdot(\dot{\xi}\rho)=0$
 on phase space (and time).
In the coordinates ($\br, \bk$), (\ref{gooddensity})
reduces to the expression (\ref{XSNdens}) of Xiao et al.,
which is hence simply the coefficient of $\prod d\xi^\alpha$ 
in (\ref{prob}), written in a local coordinate system.
The (modified) Boltzmann transport equation for $\rho$ is  
\begin{eqnarray}
\displaystyle\frac{d\rho}{dt}
&=&\nonumber
\{\rho,h\}
+\p_t\rho=
\displaystyle\frac{1}{1+e\bB\cdot\vOmega}\left[\Big(
\nabla_{\bk}\epsilon_n+e\bE\times\vOmega+e\bB(\vOmega\cdot\nabla_{\bk}
\epsilon_n)\Big)
\cdot\nabla_{\br}\rho\right.
\\[12pt]
&&\left.
-e\Big(\bE+\nabla_{\bk}\epsilon_n\times\bB
-e\vOmega(\bE\cdot\bB)\Big)\cdot
\nabla_{\bk}\rho
\right]
+\displaystyle\frac{\p \rho}{\p t}
=\displaystyle\frac{
e(\bB\cdot\vOmega)\dot{\,}}{1+e\bB\cdot\vOmega}\rho,
\label{Boltzmann}
\end{eqnarray}
where
$
(\bB\cdot\vOmega)\dot{\,}=\partial_{t}(\bB\cdot\vOmega)
+\Omega^i\dot{\br}\cdot\vnabla_{\br}{B}_i+
B_i\dot{\bk}\cdot\vnabla_{\bk}\Omega^i
$
is the material derivative. 

Using the equations of motion and
Maxwell equations $\vnabla\cdot\bB=0$ and
$\p_tB+\vnabla_{\br}\times\bE=0$, a tedious but straightforward calculation
shows, furthermore, that
\begin{equation}
\varphi=\int\!\rho d\bk
\qquad\hbox{and}\qquad
{\bf v}=\frac{1}{\varphi}\int\!\dot{\br}\rho d\bk
\end{equation}
[where the integration is over the first Brillouin zone]
satisfy the continuity equation 
$\p_t\varphi+\vnabla_{\br}\cdot({\bf v}\varphi)=0$
in {\it ordinary space-time}.
Similarly, we also have the Euler equation
\begin{equation}
\varphi\big(\p_t+\bv\cdot\vnabla_{\br}\big){\bv}=
{\bf f}-\vnabla \sigma,
\label{Euler}
\end{equation}
where ${\bf f}=\int\!\ddot{\br}\rho d\bk$ is the mean force and
$\sigma=(\sigma_{ij})$ with
$\sigma_{ij}=\int\!(\dot{x}_i-v_i)(\dot{x}_j-v_j)\rho d\bk$ 
is the kinetic stress tensor \cite{DHH}.

All these  remarks hold when
\begin{equation}
\det(\omega_{\alpha\beta})=(1+e\bB\cdot\vOmega)^2\neq0.
\label{offcritical}
\end{equation}
In the singular planar case  $eB\Omega=-1$
[which can only happen if $\Omega$ and $B$ are constants],
 the volume element (\ref{volel})
degenerates and the phase space looses $2$  dimensions.
The symplectic matrix  can not be inverted
and one can not get from a Lagrangian to the Hamiltonian framework.
The only allowed motion follow the Hall law \cite{DH,DHH,HMS}.
Hamiltonian reduction yields, however, a well-behaved 
$2$-dimensional phase space
with symplectic form proportional to the surface form in the plane, $\omega_{red}=-eBdx\wedge dy$. The fluid moves collectively, 
governed by a generalized Hall law  \cite{DHH}.


\end{document}